\documentclass[amsmath,amssymb,draftaps,showpacs,twocolumn,superscriptaddress]
{revtex4-1}
\usepackage{graphicx}% Include figure files
\usepackage{dcolumn}% Align table columns on decimal point
\usepackage{bm}% bold math
\usepackage{braket}
\usepackage{color}
\definecolor{mycolor1}{rgb}{0,0,1}
\definecolor{mycolor2}{rgb}{1,0,0}
\definecolor{mycolor3}{rgb}{0,0,0}

\newcommand{\black}[1]{\textcolor{mycolor3}{#1}}

\begin{document}

\preprint{APS/123-QED}

\title{Magnetic field effect on the chiral magnetism of noncentrosymmetric UPtGe:\\ experiment and theory}% Force line breaks with \\
%Magnetic field effect on the chiral incommensurate magnetism of UPtGe:\\ experiment and theory

\author{Atsushi Miyake}
\thanks{Corresponding author: miyake@issp.u-tokyo.ac.jp}
\affiliation{Institute for Solid State Physics, The University of Tokyo, Kashiwa, Chiba 277-8581, Japan}
 
\author{Leonid M. Sandratskii}
\thanks{Corresponding author: lsandr@mpi-halle.mpg.de}
\affiliation{Max-Planck-Institute for Microstructure Physics, Weinberg 2, Halle, D-06120, Germany}

\author{Ai Nakamura}
\affiliation{Institute for Materials Research, Tohoku University, Oarai, Ibaraki, 311-1313, Japan}

\author{Fuminori Honda}
\affiliation{Institute for Materials Research, Tohoku University, Oarai, Ibaraki, 311-1313, Japan}

\author{Yusei Shimizu}
\affiliation{Institute for Materials Research, Tohoku University, Oarai, Ibaraki, 311-1313, Japan}

\author{Dexin Li}
\affiliation{Institute for Materials Research, Tohoku University, Oarai, Ibaraki, 311-1313, Japan}

\author{Yoshiya Homma}
\affiliation{Institute for Materials Research, Tohoku University, Oarai, Ibaraki, 311-1313, Japan}

\author{Masashi Tokunaga}
\affiliation{Institute for Solid State Physics, The University of Tokyo, Kashiwa, Chiba 277-8581, Japan}

\author{Dai Aoki}
\affiliation{Institute for Materials Research, Tohoku University, Oarai, Ibaraki, 311-1313, Japan}
%\affiliation{CEA, INAC-SPSMS, F-38000, Grenoble, France}

\date{\today}% It is always \today, today,

\begin{abstract}
The effect of differently oriented magnetic field on chiral incommensurate helimagnet UPtGe is studied both experimentally and theoretically.
The magnetization measurements up to the field above the saturation have revealed an isotropic magnetic response below 20~T and a 
remarkable nonmonotonic anisotropy in high fields.
Moreover, the two principally different phase transitions from the noncollinear incommensurate to the 
%%%\blue{field-induced polarized paramagnetic state have been observed.}
\black{field-induced ferromagnetic state have been observed}.
These 
%%% apparently contradicting 
properties are successfully explained by density-functional theory calculations taking into account the noncollinearity of the magnetic structures, arbitrary directed magnetic field, and relativistic effects. 
%Interestingly, the considerable nonmonotonic anisotropy above the first order metamagnetic transition appears, although for low fields magnetic response is isotropic.
%The downward convex form of the high-field magnetization curves resembles the properties of chiral soliton lattice.
We also estimate the strength of different competing magnetic interactions and 
%%%provide plausible scenarios for the nature of the field-induced phase transformations.
\black{discuss possible scenarios of the field-induced phase transformations.}
\end{abstract}

%\pacs{ }%75.30.-m, 75.30.Kz, 75.85.+t}% PACS, the Physics and Astronomy
\maketitle

\section{Introduction} 

Uranium intermetallic compounds show a wide variety of electronic properties, owing to the delicate balance of \black{the} hybridization between 
\black{U} 5$f$ \black{electrons} and conduction electrons of ligand elements, and competing interaction energy scales \cite{Sechovsky1998}.
One of the intriguing phenomena observed in uranium based compounds is the coexistence of ferromagnetism (FM) and superconductivity \cite{Aoki2001, Huy2007, Aoki2012}.
Applying magnetic fields reveals new unexpected phenomena of these materials: e.g., 
the reentrance of the superconductivity is induced in URhGe when the magnetic field of 12~T is applied along the hard magnetization axis
%%%accompanied by the 
\black{initiating} spin-reorientation transition \cite{Levy2005}. 
Obviously, the strength of the magnetic anisotropy (MA) plays an important role in the formation of such intriguing physical phenomena. 
In contrast to the Ising-like ferromagnet URhGe, UPtGe is a unique U system showing a chiral incommensurate helical magnetic ordering below $T_{\rm N}\sim$~51~K \cite{Mannix2000}. 
%%% MA in the plane of the helix was considered to be negligibly small \cite{Sandratskii2001}.
In the model of the helix suggested in Ref.~\cite{Sandratskii2001} the MA was considered to be negligibly small.
The spin dynamics was recently studied via the NMR experiments, and the XY-type spin fluctuations were clarified \cite{Tokunaga2018}.
The chiral helical structures in various types of materials have been attracting strong research interest because of 
their importance in the physics of skyrmion lattices and chiral domains, topics of intense study
in the field of spintronics \cite{Muhlbauer2009,Heinze2011,Bergmann2014}.
%%%their relation to the questions 
%%%in the field of spintronics, for example, skyrmion lattices and chiral domain \cite{Muhlbauer2009,Heinze2011,Bergmann2014}. 

\begin{figure}[t]
   \begin{center}
      \includegraphics[width=65mm]{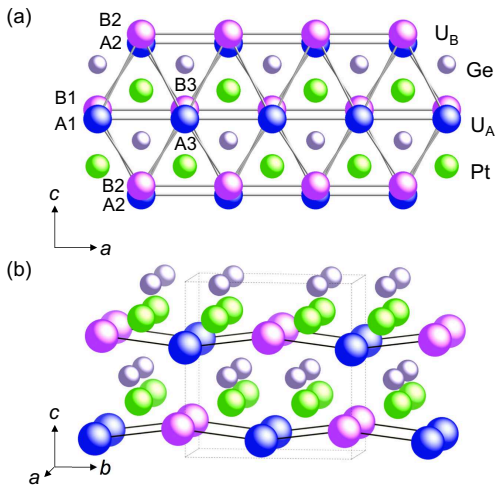}
   \end{center}
   \caption{Crystal structure of UPtGe projected on the (a) $ac$ and (b) $bc$ planes. 
There are two inequivalent U sublattices referring U$_{\rm A}$ and U$_{\rm B}$, whose closest neighbors are Pt and Ge atoms, respectively \cite{Hoffmann2001}. 
The U atoms lying in the same $ac$ plane form a distorted hexagonal lattice formed by the atoms of both sublattices. 
The labels of U atoms in (a) are used in Table~\ref{table1}.
Along the $b$ axis the U atoms of two sublattices form zigzag chains.
}
   \label{structure}
\end{figure}

A number of neutron diffraction studies lead to the same conclusion that the magnetic ground state of UPtGe is an incommensurate cycloid propagating along the $a$ 
axis with the wave vector $\vec{q}=[0.55-0.57, 0, 0]$ in units of $2\pi/a$; the U moments lie  in the $ac$ plane \cite{Szytula1992, Kawamata1992, Robinson1993, Mannix2000}.
Importantly, Mannix {\it et al}. \cite{Mannix2000} clarified that the orthorhombic crystal structure of UPtGe is of the noncentrosymmetric EuAuGe type ($Imm2$, space group No. 44) 
%% LS we can save here some space
%%(space group $Imm2$) 
(Fig.~\ref{structure}), which is different from the centrosymmetric TiNiSi type, such as of URhGe, or CeCu$_2$ type assumed in previous studies \cite{Szytula1992, Kawamata1992, Robinson1993,Hoffmann2001,comment_lattice}. 
%It is noted that the $a$ and $b$ axes are switched from the TiNiSi to EuAuGe structure.
%%%Another revealed important property 
\black{Another important property revealed in the experiment}
is a chiral character of the cycloid \cite{Mannix2000}: All domains have the same wave vector $\vec{q}$ whereas the domains with opposite wave vector $-\vec{q}$ 
%%%were 
\black{are} absent. 

Previously proposed explanation \black{\cite{Sandratskii2001}} of the origin of the incommensurate cycloid in UPtGe includes 
the following components:
(i) a very small MA in the cycloid plane, which was treated as negligible, (ii) competing interatomic exchange interactions, and (iii) an active Dzyaloshinskii-Moriya interaction (DMI) 
due to the lack of the inversion symmetry. 
%%%\cite{Sandratskii2001}. 
The first two features explain the formation of the incommensurate cycloid, and the third is responsible for the chiral character of the magnetic ground state. 
An important conclusion of Ref.~\cite{Sandratskii2001} is a critical sensitivity of the magnetic structure to the crystal lattice \cite{comment_lattice_sens}. 

The study of 
%%%magnetic field response on 
the response of UPtGe to the magnetic field
is expected to deepen the understanding of chiral incommensurate magnetism and of the field induced 
incommensurate-commensurate (I-C) phase transitions, 
which are 
%%%an intriguing phenomenon 
fascinating phenomena of solid state physics. 
%Early reviews on this topic are already more than 30 years old . 
Despite the long history of 
the studies on this topic (see, e.g, early reviews \cite{Bak82,Izyumov1984})
%%% magnetic studies more than 30 years \cite{Bak82,Izyumov1984}, 
the understanding of the I-C transitions is by far not complete. 
There are some exact 
%%%theoretical 
statements, which are 
%%%merely 
based on very simple theoretical models, like sine Gordon equation \cite{Dzyaloshinskii1964,Togawa2016}. 
%%%but 
The applicability of these models to complex real materials is not self-evident. 
There are also more complex theoretical models solved numerically, e.g. atomistic spin Hamiltonians \cite{Jensen1996}. 
However, such models are sensitive to the values of a large number of parameters whose choice is not unique.

We report a joint experimental and theoretical study of the magnetic field effect on UPtGe. 
The magnetization measurements are performed in fields up to 56~T that are above the saturation field 
for field directions in the cycloid plane.
%%%with field directions in the $ac$ plane, using 
The 
%%%observed 
isotropic behavior 
observed  in low fields is replaced at higher fields by anisotropic field-induced phase 
transitions. 
This raises new important questions that we address on the basis of the density-functional theory (DFT) calculations.  
One of the focuses of our attention is the complex interplay of various interactions responsible for the unusual physical properties of UPtGe. 
Other focus is the field-induced phase transitions.
%%Another focus is the field-induced phase transitions.
Our 
%%%experiments reveal 
\black{experiment reveals} the sequence of two very different phase transformations bringing the system from the chiral incommensurate 
helical state to the field-induced 
%%%\blue{polarized paramagnetic (PPM)} 
\black{ferromagnetic state  \cite{comment_field_induced_FM}}. 

\begin{figure}[t]
   \begin{center}
      \includegraphics[width=85mm]{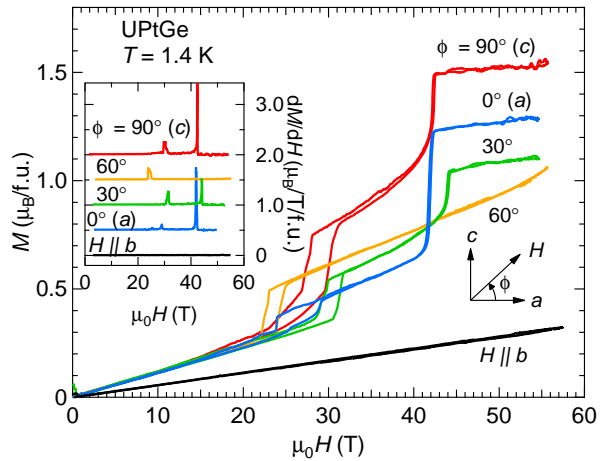}
   \end{center}
   \caption{Magnetization $M(H)$ curves of UPtGe at $T=1.4$~K for various field directions.
   The inset shows the differential susceptibility d$M$/d$H$ for each increasing-field sweep.
   d$M$/d$H$ are offset by 0.5~$\mu_{\rm B}$/T/f.u. for the sake of clarity. 
% $\phi$ is the angle between the fields and the $a$ axis in the $ac$ plane.
   }
   \label{experiment}
\end{figure}

\section{Experiment} 
Single crystals of UPtGe were prepared using the Czochralski pulling method in a tetra-arc furnace.
Pulsed magnetic fields were generated using nondestructive magnets installed at the International MegaGauss Science Laboratory of the Institute for Solid State Physics at the University of Tokyo.
The conventional induction method using coaxial pick-up coils was used for the magnetization measurements down to 1.4~K.

\section{Results of experiment}
Magnetization $M(H)$ curves 
%%%along 
\black{for} various field directions are presented in Fig.~\ref{experiment}. 
%%AM : Shortened for make space.
%%Our experimental results for the magnetization in the fields with varying spatial directions are presented in Fig.~\ref{experiment}. 
The anisotropy between the $ac$ plane and out-of plane $b$ axis is very large.
For $H~||~b$, the magnetization is $H$ linear without any anomalies and much smaller than 
%%%those 
\black{for the field} in the $ac$ plane.
%%Here
In the following, we focus on the field directions in the cycloidal $ac$ plane.
%%%The field directions $\phi$ 
\black{Angle $\phi$ defining field direction} is measured from the orthorhombic $a$ axis.
Below 20~T, the magnetization is \black{nearly} isotropic.
This isotropic magnetic response seems to be 
%%%that 
expected in connection with the assumed negligibly small in-plane MA \cite{Sandratskii2001}. 
By contrast, two remarkable anisotropic increments of magnetization are seen at  about 25~T and 42~T (Fig.~\ref{experiment})
%%%, 
revealing the presence of considerable MA in the $ac$ plane. 
The observed high-field anisotropy is nonmonotonic with respect to the field direction.
Indeed, the saturated magnetic moment is maximal for the field parallel to the $c$ axis ($\phi=90^\circ$). 
It strongly drops for $\phi =60^\circ$, and then increases again for $\phi =30^\circ$ and $\phi =0^\circ$ ($a$ axis) (Fig.~\ref{experiment}).
%%%Interestingly, 
\black{F}or $\phi =60^\circ$, magnetization shows upturn near the maximum fields but does not saturate up to the maximum fields. 
%%% and at 1.4~K. 
%, indicating that the hardest in-plane axis is not along the orthorhombic principal axes.

\begin{figure}[t]
\begin{center}
\includegraphics[width=80mm]{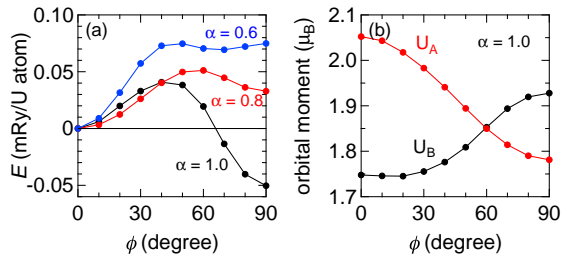}
\end{center}
   \caption{(color online) (a) Magnetic anisotropy energy (MAE) for the FM configurations and (b) orbital moments of the two U sublattices as a function of  $\phi$.
%(a) Magnetic anisotropy energy for the ferromagnetic configurations as a function of the direction of the magnetic moments
%in the $ac$ plane. 
%(b) Orbital moments of the two U sublattices as functions of $\phi$.
}
   \label{FIG_MAE_UPtGe_3curves.eps}
\end{figure}

\section{Method of calculation} 
To understand the 
%%%above 
unusual angular dependence of the in-plane MA, we performed the calculation of the energy of the FM configuration for different directions of the atomic moments with respect to the crystallographic axes.
The calculations were carried out with the augmented spherical waves code \cite{Williams1979,Eyert2012}
%%%, being
able to deal with noncollinear magnetism, spin-orbit coupling, and magnetic field along an arbitrary \black{direction} \cite{Sandratskii1998,Sandratskii2016}. 
Since the localization of the 5$f$ states can be underestimated in standard DFT calculations, we performed additional calculations introducing a scaling parameter $\alpha<1$ to study the sensitivity of the results to the level of the 5$f$-states localization. 
This parameter is used as a scaling factor for interatomic Hamiltonian and overlap integrals of U 5$f$ wave functions entering the secular matrix.
The exchange-correlation potential was used in the local density approximation (LDA) \cite{Barth1972}. 
The {\bf k}-vector sampling suggested by Monkhorst and Pack \cite{Monkhorst1976}  was employed in the integration over 
the Brillouin zone (BZ). 
The description of the crystal structure of the EuAuGe type
and lattice parameters of UPtGe determined in the neutron diffraction experiment can be found in Ref.~\cite{Mannix2000}.  
In the calculations for orthorhombic unit cell containing four 
U atoms the number of the {\bf k} points in the BZ varied between 8000 and 27000.
For larger unit cells the number of the {\bf k} points decreased in accordance with 
decreasing BZ volume. In very long-lasting calculations of self-consistent magnetic 
structures in large supercells the reduced numbers of the k points were used.

\section{Results of calculations and discussion} 
A prominent feature of the theoretical $\phi$ dependence of the magnetic anisotropy energy, $E(\phi$) [Fig.~\ref{FIG_MAE_UPtGe_3curves.eps}(a)] is its nonmonotonic character, which is consistent with the $M(H,\phi)$ curves  (Fig.~\ref{experiment}).  
An insight into the origin of the nonmonotonic $E(\phi$) is provided by the consideration of the $\phi$ dependence of the orbital moments of the two inequivalent U sublattices [see Fig.~\ref{FIG_MAE_UPtGe_3curves.eps}(b)],
since there is deep physical connection between MA energy and orbital moments anisotropy (see, e.g., Refs.~\cite{Bruno1989,Sandratskii2015}).
%% I suggest new version above. 
%%\blue{The anisotropy of the total moment is usually mostly from the orbital moment rather than the spin moment (see, e.g., Refs.~\cite{Bruno1989,Sandratskii2015}).}
%The correlation of the anisotropy of the orbital moments and MAE was many times discussed in the literature (see, e.g., Refs.~\cite{Bruno1989,Sandratskii2015}).
The calculations gave the remarkable result that the orbital moments of the two U sublattices, though both monotonic functions of $\phi$, have opposite character:
decrease for the A sublattice and increase for the B sublattice.
%%% as a function of $\phi$.
The competition of two opposite angular dependence\black{s} explains unexpected properties of the MA of UPtGe: its \black{unusual smallness for U compounds} and the nonmonotonic behavior.
The calculations with reduced overlap of the 5$f$ functions scaled with parameter $\alpha=0.8$ show that the nonmonotonic behavior of the $E(\phi$) is a robust property [Fig.~\ref{FIG_MAE_UPtGe_3curves.eps}(a)]. 
For the stronger scaling with parameter $\alpha=0.6$, the nonmonotonic features become weak. 

\begin{figure}[t]
  \begin{center}
      \includegraphics[width=85mm,clip=true]{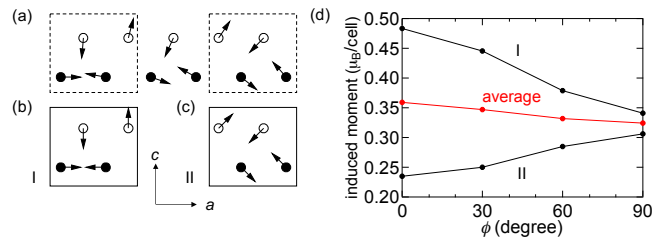}%{spiral_uptge.eps}
   \end{center}
   \caption{(a) Schematic picture of an incommensurate cycloid with wave vector $q=0.527$.
The open (full) symbols correspond to U$_{\rm A}$ (U$_{\rm B}$), respectively.
    (b) Supercell I corresponding to $q$=0.5 and atomic moments parallel to the $a$ and $c$ axes.
    (c) Supercell II: atomic moments rotated by $45^\circ$ with respect to supercell I. 
    (d) The induced magnetic moments of supercells I, II, and their average as a function of $\phi$.
\black{The supercells presented in (b) and (c) have an increased size of 2$a$ along the $a$ axis. 
Only the atoms laying in the same $ac$ plane are shown.}
}
   \label{spiral_uptge.eps}
\end{figure}

After we have revealed the presence of sizable in-plane MA 
%%%the isotropic response below 20~T (Fig.\ref{experiment}) appears rather contradictory. 
for the FM structures, we would like to reveal the origin of the isotropic response
below 20~T (Fig.\ref{experiment}).
A direct DFT-calculation of the response of an incommensurate magnetic structure to differently oriented magnetic field is not feasible because of an infinite magnetic unit cell of the incommensurate structure.
Therefore, we performed the following calculations to address this problem. 
We selected two parts of the helix with different directions of the atomic moments with respect to the crystal lattice [see Fig.~\ref{spiral_uptge.eps}(a)] 
and described these pieces with the supercells corresponding to commensurate helix with $q=0.5$.
In the first supercell, the initial directions of the atomic moments are collinear to the $a$ and $c$ axes  [Fig.~\ref{spiral_uptge.eps}(b)]. 
In the second, the moments were rotated by 45$^\circ$ [Fig.~\ref{spiral_uptge.eps}(c)]. 
Next we calculate the magnetic response 
%%%in a 
\black{to the} field of 23.5~T \cite{comment_23T} applied along different directions.
We obtained two opposite monotonic dependencies of the induced moment on the field direction $\phi$ for two supercells [Fig.~\ref{spiral_uptge.eps}(d)].
This property explains the isotropy of the response of the cycloid to the applied magnetic field below 20~T as the result of the averaging of the anisotropic responses of different parts of the cycloid. 

\black{It is worth noting that the contributions of the orbital and spin moments to the induced moment shown in Fig.~\ref{spiral_uptge.eps}(d) have opposite
signs: positive for the orbital moment and negative for the spin moment. The ratio of the magnitudes of the induced orbital and spin moments varies for the
points of the `average' line in the interval between 2.2 and 2.4. The fact that the induced spin moment is opposite
to the direction of the magnetic field reveals stronger influence of the third Hund's rule than the direct influence of the Zeeman coupling to the field
(see also Ref.~\cite{Sandratskii2016}).}  

\begin{figure}[t]
   \begin{center}
      \includegraphics[width=65mm]{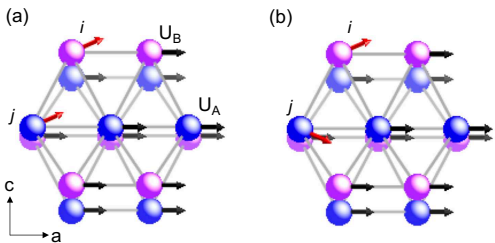}
   \end{center}
   \caption{The difference of the energies of the two magnetic configurations (a) and (b) gives an estimation
of the exchange interaction parameter $J_{ij}$ between atoms $i$ and $j$.
}
   \label{schematic}
\end{figure}

It is important to compare the energy scales of different magnetic interactions, i.e., interatomic exchange interaction, DMI, and Zeeman energy.
First, we estimate the interatomic exchange parameters. 
As a reference state of the system, we used the FM configuration with atomic moments parallel to the $a$ axis.
To estimate the exchange interaction parameter between atoms $i$ and $j$ in Fig.~\ref{structure}(a), 
we evaluated the energies of the magnetic configurations with the moments of atoms $i$ and $j$ deviated 
in the $ac$ plane by angle 10$^\circ$ in the same and opposite directions \black{(see Fig.~\ref{schematic})}.
%%%\red{*** To the caption: The difference of the energies of the two magnetic configurations gives an estimation of the exchange interaction between atoms $i$ and $j$.***} 
The difference of these energies estimates the exchange energy corresponding to the angle 20$^\circ$ between moments of the atoms $i$ and $j$.
By dividing the energy by $[1-\cos{(20^\circ)}]$, we obtain exchange parameter $J_{ij}$. 
The values of the calculated exchange parameters 
are listed in Table~\ref{table1}.

%\begin{figure}[t]
%  \begin{center}
%      \includegraphics[width=85mm,clip=true]{exchange_calc_scheme}
%   \end{center}
%   \caption{The difference of the energies of the two magnetic configurations gives an estimation
%of the exchange interaction between atoms $i$ and $j$.}
%   \label{exchange}
%\end{figure}

\begin{table}[h]
\caption{Interatomic exchange parameters $J_{\it ij}$ in the units of mRy.}

\begin{center}
\begin{tabular}{c c c c }

\hline
 $i$~,~$j$  & $\alpha=1.0$ ~~& $\alpha=0.8$ ~~& $\alpha=0.6$ \\ 
% \\ 
\hline
A1~,~B1      & 0.275  &  0.821 & 1.373  \\
A1~,~A2   &  -0.013 &  -0.014 &   -0.014 \\
 A1~,~B2   & -0.076  &  0.009  & 0.139\\
 A1~,~A3      &  -0.202 &  -0.239  &  0.061 \\
A1~,~B3 &  -0.047 &    -0.105 & -0.123\\
B1~,~B2 &   -0.058 &  -0.017 & -0.0257\\
B1~,~B3 &  -0.014 &  0.061 & 0.082\\
\hline

\label{table1}
\end{tabular}
\end{center}
\end{table}

For unscaled calculation ($\alpha=1.0$), only exchange parameter $J_{\rm A1, B1}$ is FM. 
\black{It corresponds to the interaction between neighboring atoms of the zigzag chain (Fig.~\ref{structure}).
This interaction is the strongest among all estimated interatomic exchange interactions.} 
%%%corresponding to the strongest interaction between neighboring atoms of the zigzag chain (Fig.~\ref{structure}), and is the strongest.
%% of all estimated interatomic exchange interactions. 
This result supports the picture of ferromagnetically ordered zigzag chains  \cite{Kawamata1992,Robinson1993, Mannix2000}. 
%%AM: To save space, the sentence is shortened.
%%This result provides the support for the picture of ferromagnetically ordered zigzag chains \red{proposed} previously \cite{Kawamata1992,Robinson1993, Mannix2000}. 
The decrease of \black{the} overlap \black{given by $\alpha<1$}, on one hand, increases the values of atomic moments and, on the other hand, diminishes the quantum-mechanical interaction integrals. 
Therefore, there is no $a~priori$ way to relate the change in the electronic overlap to the character of the variation of interatomic exchange interaction.
Indeed, the analysis of exchange parameters presented in Table~\ref{table1} shows that there is no general trend in the variation of the exchange interactions with decreasing scaling parameter.
For the calculation without scaling, all exchange parameters with the exception of the interchain one are antiferromagnetic (AFM) that leads to the frustration and canting of the atomic moments. 
The strongest AFM interaction is $J_{\rm A1, A3}$ [Fig.\ref{structure}(a)], while $J_{\rm B1, B3}$ is remarkably weak. 
%%AM: shortened
%%The strongest interaction is between atoms 1 and 3 of the A sublattice [Fig.\ref{structure}(a)]. 
%%Remarkably, the interaction between atoms 1 and 3 of the B sublattice is very weak. 
For scaling factor $\alpha=0.8$ the values of parameters are in good correlation with those for unscaled calculations. 
For stronger scaling with $\alpha=0.6$ the exchange interactions become distinctly more FM.

%One coauthor suggested to add sentence(s) why we evaluate the strength of DMI and Zeeman energy. Please modify.
The chiral magnetism of UPtGe essentially arises
%%is arisen 
from the DMI \cite{Sandratskii2001}.
%%LS: to save space
%%It is important to compare with the other energy scales.} 
The strength of DMI is calculated as the difference of the energies of the cycloids with $q$=$0.5$ and $q$=$-0.5$ and 
%%%is obtained as follows, 
\black{has the following values:} 0.11~mRy/U for unscaled calculation, 0.21~mRy/U for $\alpha=0.8$ and 0.22~mRy/U for $\alpha=0.6$. 
%%The strength of DMI is calculated as the difference of the energies of the cycloids with $q$=$0.5$ and $q$=$-0.5$. 
%%We obtained the following values of the DMI parameter per U atom: $D=0.11$~mRy for unscaled calculation, $D=0.21$~mRy for scaling with $\alpha=0.8$ and $D=0.22$~mRy for scaling with $\alpha=0.6$. 
This estimation shows
%%%, importantly, 
that the energy scale of the DMI is of the same order of magnitude as the exchange interaction. 
We also obtained considerable dependence of the DMI strength on the localization parameter. 

Finally, we estimate the scale of the Zeeman energy.  
The orientation of atomic moments parallel to the magnetic field gives the energy gain of $\mu_0HM$ where $M$ is the value of the atomic magnetic moment per U atom. 
At 40~T and $M= 1.25~\mu_{\rm B}$ we obtain the value of $\mu_0HM\sim$~0.21 mRy/U, which is close to the values of the AFM exchange interactions (see Table~\ref{table1}). 

%%%As seen above, we have found 
\black{As shown above the calculations give}
a complex balance of 
%%%the 
several interactions.
%% having similar orders of magnitude. 
The chiral magnetic ground state of UPtGe comes from competing exchange interactions and contributions of the DMI and MA. 
In applied magnetic fields, the Zeeman energy overcomes these interactions, leading to the FM transformation. 

Now we turn to the discussion of the two field-induced phase transitions (Fig.~\ref{experiment}). 
The applicability of the direct DFT calculations to the description of these phase transitions is rather limited \cite{comment_limitation}.
%%AM: To save the space, the explanation of the limitation is better to move to reference place. You agree?  
%%One of the limitations is the necessity to use large supercells. 
%%Another limitation is the absence in the standard DFT calculations of the fluctuations responsible in the nature for the realization of the first order phase transitions.   
Nevertheless, it is instructive to perform the following calculations.
We consider supercells of different moderate sizes and start the iterational process with various accidentally chosen magnetic configurations. 
The magnetic moments are allowed to relax to a self-consistent state.
On the next step, we apply the magnetic field and consider the change of the magnetic configurations. 
The calculations with self-consistently determined directions of the magnetic moments were 
performed as follows. 
In the first step, the calculations were carried out with a relatively small 
number of $k$ points in the BZ that varied from about 200 for the supercell containing eight U atoms to about 50
for the supercell with 16 U atoms. When the convergence of the directions of all U moments
reached 0.1$^\circ$ the number of $k$ points was approximately doubled and the calculations continued until 
the convergence of the directions is better than  0.01$^\circ$ \cite{comment_involvedCalc}. 

\begin{figure}[t]
   \begin{center}
      \includegraphics[width=80mm]{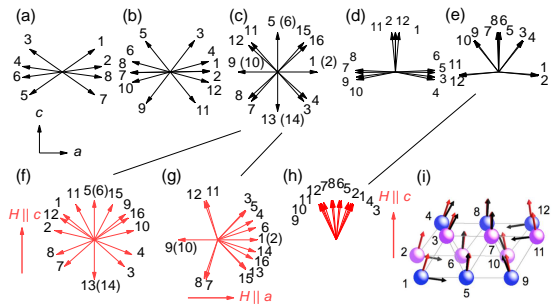}%{FIG_magn_struct.eps}
   \end{center}
   \caption{Examples of self-consistent magnetic structures. (a)-(e) Magnetic structures without applied magnetic field. 
Magnetic structures (f) for $H~||~c$ and (g) for $H~||~a$ based on the magnetic structure (c).
(h) Magnetic structure for $H~||~c$ based on the magnetic structure (e).
(i) Space resolved presentation of structures (e) and (h) with correspondingly black and red vectors.
The numbering of atoms is shown in (i).
The number of the U atom in the supercell is 8 for structure (a), 12 for structures (b), (d), (e), and (h), 
and 16 for structures (c), (f), and (g). The supercells were obtained by extending the unit cell along the $a$ axis.
} \label{FIG_magn_struct.eps}
\end{figure}

The analysis of performed calculations shows that starting from different initial magnetic configurations we generally obtain different self-consistent magnetic states. 
Among the self-consistent states there are both magnetically compensated [see examples in Figs.~\ref{FIG_magn_struct.eps}(a), \ref{FIG_magn_struct.eps}(b), and \ref{FIG_magn_struct.eps}(c)] and uncompensated states [Figs.~\ref{FIG_magn_struct.eps}(d) and \ref{FIG_magn_struct.eps}(e)].
The fact that the calculations give multiple self-consistent states indicates the presence of numerous local minima in the complex high-dimensional energy landscape describing the energy of the system as a function of the directions of atomic moments.

Figures~\ref{FIG_magn_struct.eps}(f), \ref{FIG_magn_struct.eps}(g), and \ref{FIG_magn_struct.eps}(h) show the result of the calculations with external field of 100 T.
In Figs.~\ref{FIG_magn_struct.eps}(f) and \ref{FIG_magn_struct.eps}(g), we present the transformation of the compensated magnetic structure (c) 
in the fields directed along the $c$ and $a$ axes, respectively.
For both field directions there is the modification of the magnetic structure resulting in an induced magnetic moment.
The response to the magnetic field is anisotropic. 
The induced net spin, orbital and total magnetic moments along field direction for structure (f) are, respectively, $-3.68$, 7.60, and 3.92 $\mu_{\rm B}$ per supercell. 
For structure (g) these values are larger: $-6.92$, 12.37, and 5.45~$\mu_{\rm B}$.
%$-1.13$, 3.50 \red{and} 2.37  $\mu_{\rm B}$. 
The energies of the in-field structures (f) and (g) are lower than the energy of the compensated structure (c) by 6.2 and 3.1~mRy, respectively. 
The decreases in the Zeeman energy are, however, only 2.0 and 2.7~mRy per supercell. 
This reveals that self-consistent response of the electron system to the applied field is considerably more complex than a rotation of the rigid atomic moments.

Figure~\ref{FIG_magn_struct.eps}(h) shows the transformation of the uncompensated magnetic structure (e) with the field along the $c$ axis. 
Interestingly, in this case the modification of the magnetic structure is especially strong and the in-field state is rather close to the FM state. 
The numerical results give the following values: the differences of the net spin, orbital, and total moments of the structures (e) and (h) are $-9.3$, 16.3, and 7.0 $\mu_{\rm B}$
per supercell.
The energy of structure (h) is lower than that of structure (e) by 5.9 mRy, whereas the difference of the Zeeman energies of the two structures is only 3.5 mRy per supercell.

We remark that the magnetic field used in the calculations presented in Figs.~\ref{FIG_magn_struct.eps}(f), (g) and (h) is about two times larger than the maximal experimental field. 
This large field was selected to make the influence of the field clearly visible in the figures. 
The property that this field does not lead in all cases to the magnetic structure close to ferromagnet is explained as follows. 
The energy of the system as a function of the 
directions of the atomic moments is a very complex unknown function with many local minima and barriers separating these minima. 
It is to be expected that these barriers are often higher than the characteristic Zeeman energies even for large magnetic fields.
In the DFT calculations we deal with electron interactions that are of larger scale. 
Since in the calculations there are no fluctuations that can be responsible for the tunneling of the system from a local minimum to a deeper minimum the system 
remains near one of the quasistable states. 
It is also important that we can perform calculations only for relatively 
small supercells whereas in the I-C transitions the intermediate structures with larger period can be important.
This makes 
the problem of the quantitative theoretical description of the system very challenging. 
Our paper makes a step towards this 
description but remains on the level revealing qualitatively new properties. 
   
Coming back to our experiment, we remark that the interpretation of the lower-field phase transition at around 25-30~T is rather straightforward.
The system transforms discontinuously in a magnetically uncompensated fan-type structure. 
The examples of such a structure are shown in Figs.~\ref{FIG_magn_struct.eps}(d) and \ref{FIG_magn_struct.eps}(e) \cite{comment_fan_structure}.

The nature of the higher-field phase transition to
\black{the ferromagnetic state} 
around 40~T is principally different, as clearly seen in Fig.~\ref{experiment}. 
First, it does not have a noticeable hysteresis. 
Second, the $M(H)$ curves just below the transition have a strong \black{convex} curvature. 
There are two possible scenarios of this transition.
One 
possibility is a special property of the energy landscape of UPtGe consisting in 
(i) almost equal energies of the fan structure just before the transition and the 
\black{ferromagnetic} structure after 
the transition and (ii) the existence of a barrier-free path between these points of the energy landscape.
The process resembling this type of transformation is obtained in our numerical experiments [Figs.~\ref{FIG_magn_struct.eps}(e),\ref{FIG_magn_struct.eps}(h), and \ref{FIG_magn_struct.eps}(i)]. 
This calculation shows that the system can, in principle, relatively easily move from the fan configuration to the state close to collinear ferromagnetism. 

As an alternative scenario, we 
%%%briefly 
mention the possibility of soliton-lattice formation.
Here, with increasing fields the regions of noncollinear magnetic moments are separated by increasing ferromagnetically aligned domains (see, e.g., figures in Refs.~\cite{Zheludev1998,Togawa2016} illustrating this kind of transition).
The \black{convex} $M(H)$ curve and hysteresis-free transitions are characteristic features of the soliton-lattice type transition \cite{Togawa2016}
%%%, 
which is consistent with the in-field behavior obtained 
%%%from 
\black{as} the solution of the sine-Gordon equation.

It is possible that both scenarios contribute to the transition.
The neutron diffraction or resonant x-ray scattering studies of the transitions would be very useful but are challenging in such high-field regions.
On the theoretical side, one can study the energetics of the system using a lattice spin-model. 
However, a very large crystal domain must be used to describe the structures of different periodicity, and a large number of parameters must be employed to reflect the complexity of the system. 
Though further progress appears rather laborious, we hope that our work will 
stimulate new deep studies on \black{the} I-C transition of the chiral helical magnetic structures. 

\section{Summary} 
We have reported the magnetic properties of chiral incommensurate magnet UPtGe in the fields of varying directions and up to above the saturation.
We have revealed that magnetic response, isotropic for fields below 20~T, becomes strongly anisotropic for higher fields where two principally different phase transitions are observed.
Remarkably, this anisotropy possesses an unusual nonmonotonic field-orientation dependence.
%is nonmonotonic with respect to the direction of the magnetic field. 
Our DFT calculations successfully explain the \black{apparently} contradicting properties 
%%%between low- and high-fields 
\black{obtained in the low- and high-field experiments}
with the identification of competing contributions into magnetic interactions, magnetic anisotropy, and magnetization process.  
We suggest an interpretation of the nature of the two phase transitions that bring the system from the incommensurate to the 
\black{field-induced ferromagnetic state}.
%%%\blue{field-induced PPM phase}. 
Our work deepens the understanding of the physical origin of the wide variety of the properties of the U intermetallics.
%Our work has now clarified the whole picture of complex magnetic properties in the noncentrosymmetric uranium intermetallic system with hybridization effects. 

\begin{acknowledgments}
The authors are grateful to A. Pourret and Y. Tokunaga for fruitful discussions.
This research was carried out (in part) at the International Research Center for Nuclear Materials Science, 
Institute for Materials Research, Tohoku University.
This work was partially supported by the MEXT of Japan Grants-in-Aid for Scientific Research (JP15K17700, JP15K05156, JP15H05882, JP15K05884, JP15K21732, JP15KK0149, and JP16H04006).
\end{acknowledgments}

%\bibitem{Aoki2006} D. Aoki, T. D. Matsuda, V. Taufour, E. Hassinger, G. Knebel, and J. Flouquet, J. Phys. Soc. Jpn. {\bf 78}, 113709 (2009).

%\bibitem{Robinson1994} R. A. Robinson, A. C. Lawson, V. Sechovsky, L. Havela, Y. Kergadallan, H. Nakotte, and F. R. de Boer, J. Allots. Comp. {\rm 213-214}, 528 (1994).

\end{document}